# Anisotropic thermal expansion and electronic transitions in the $Co_3BO_5$ ludwigite

N.V. Kazak[1], A. Arauzo[2,3], J. Bartolomé[2], M.S. Molokeev[1,4], V.A. Dudnikov[1], L.A. Solovyov[5], A.A.Borus[1], S.G. Ovchinnikov[1]

[1]*Kirensky Institute of Physics, Federal Research Center KSC SB RAS, 660036 Krasnoyarsk, Russia*
[2]*Instituto de Nanociencia y Materiales de Aragón (INMA), CSIC-Universidad de Zaragoza and Departamento de Física de la Materia Condensada, 50009 Zaragoza, Spain*
[3]*Servicio de Medidas Físicas, Universidad de Zaragoza, 50009 Zaragoza, Spain*
[4]*Research and Development Department, Kemerovo State University, Kemerovo, 650000, Russia*
[5]*Institute of Chemistry and Chemical Technology, Federal Research Center KSC SB RAS, 660036 Krasnoyarsk, Russia*

The investigations of the crystal structure, magnetic and electronic properties of the $Co_3BO_5$ at high temperatures were carried out using powder x-ray diffraction, magnetic susceptibility, electrical resistivity, and thermopower measurements. The orthorhombic symmetry (Sp.gr. *Pbam*) was established at 300 K and no evidence of structural phase transitions was found up to 1000 K. The thermal expansion of the crystal lattice is strongly anisotropic. At $T<T_C$=550 K, a large thermal expansion along the *c*-axis is observed with simultaneous contraction along *a*-axis. The activation energy of the conductivity decreases significantly at high temperatures and follows the thermal expansion variation, that exhibits two electronic transitions at ~500 and ~700 K, in coincidence with the anomalies of the heat capacity. Electronic transport was found to be a dominant conduction mechanism in the entire temperature range. The temperature dependence of the effective magnetic moment reflects the evolution of the spin state of $Co^{3+}$ ions towards the spin crossover to a high spin state. The interrelation between the crystal structure and electronic properties is discussed.

## I. INTRODUCTION

The oxyborates with ludwigite structure are of current interest due to an intriguing magnetic behavior, electronic and structural transitions, and strong magnetic anisotropy [1-6]. The chemical formula $Me_2^{2+}Me^{3+}BO_5$ contains two divalent and one trivalent metal ions, which can be represented by the *3d* metal, $Mg^{2+}$, $Al^{3+}$, and $Ga^{3+}$. The ludwigite structure permits a wide row of isomorphous substitutions, which allows the study of successive transformations of the physical properties, from which the magnetic ones are studied most extensively [7-11].

Homometallic ludwigites take a special place due to the metal ions $Me^{2+} = Me^{3+}$ occupy the sites of the same symmetry with an interionic distance of less than 3 Å, thereby providing conditions for charge-ordering. The $Co_3BO_5$ is one of the known homometallic ludwigites, which has been intensively investigated last years. The compound has an orthorhombic symmetry with space group *Pbam*(No. 55). The crystal structure consists of a network of edge-sharing octahedra and planar $BO_3$ triangles and contains four nonequivalent positions for cations: M1(2a), M2 (2b), M3 (4g), and M4 (4h), five positions for oxygen, and one position for boron atoms (Fig. 1). The divalent cobalt ions prefer to occupy the M1, M2, and M3 sites, while the M4 site is filled by trivalent cobalt ions as it was found from x-ray and neutron powder diffractions measurements [12,13]. The $Co_3BO_5$ displays a ferrimagnetic transition at $T_N$=43 K with a *b*-axis as an easy magnetization direction and no structural phase transition was found on cooling down to 2 K [12,18,19].

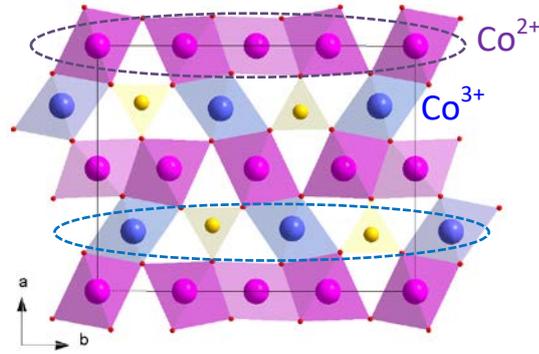

Fig. 1. The crystal structure of the $Co_3BO_5$ presented in the *ab*-plane. The octahedra occupied by divalent and trivalent cobalt ions are highlighted by pink and blue, respectively. The planar $BO_3$ groups are shown by the yellow triangles. The dotted lines show the layers built of the divalent and trivalent cations.

The magnetic and electronic states of the Co ions were thoroughly studied in a wide temperature range including the magnetically ordered and paramagnetic phases using different experimental techniques [12,13,20,21]. The neutron powder diffraction, x-ray magnetic circular dichroism at Co K-edge, and paramagnetic susceptibility measurements revealed that below room temperature the contribution to the magnetic behavior of $Co_3BO_5$ is due to the high-spin $Co^{2+}$ ions, while the contribution of the $Co^{3+}$ ions is suppressed due to its low-spin state. The high-temperature x-ray diffraction measurements on the $Co_3BO_5$ single crystal revealed a sharp increase in the bond-lengths at the M4 site at ~500 K which was attributed to the transition from low-spin to high spin state of $Co^{3+}$ ion and originates from the difference in the ionic radii of $Co^{3+}$ ion ($r_i$=0.545 Å in the low-spin and 0.61 Å in the high-spin states) [22]. This electronic transition correlates well with the observation of an anomaly of the thermal expansion and heat capacity at 500 K. A second electronic transition at ~700 K was observed from heat capacity measurements, the origin of which requires an additional study.

The theoretical support of the spin-state transition was obtained through the bond structure calculations using the GGA + U approach [12]. In the low-temperature phase, the $Co_3BO_5$ is characterized by a charge-ordering with the Co $d^6$ ions filling the M4 site and $d^7$ ions located at M1, M2, and M3 sites. This phase corresponds to the M1(↓)-M2(↑)-M3(↑)-M4(0) spin configuration, where ↑ stands for a spin up and ↓ for spin down, being the $Co^{2+}$ ions in a high-spin state and the $Co^{3+}$ ions nonmagnetic. An insulating ground state was found to be related to the charge-ordering. At high temperatures, the breaking of the charge-ordering occurs resulting in the formation of a metallic state. The GGA + U calculations show that at high temperatures all Co ions including $Co^{3+}$ reach the high-spin state, clearly indicating that the $Co_3BO_5$ undergoes spin-state transition on heating.

Thus, the investigation of the cobalt ludwigite is gradually shifting to the high-temperature region with an emphasis on the lattice dynamics and the electronic transitions caused by the spin transformation and, possibly, charge delocalization. The lattice dynamics relevant to anomalous thermal expansion of $Co_3BO_5$ was studied through x-ray diffraction, x-ray absorption (EXAFS), differential scanning calorimetry, and heat capacity measurements [12, 20, 21]. Although the high-temperature x-ray diffraction studies of the $Co_3BO_5$ performed both on the single-crystalline and powder samples did not reveal structural phase transitions up to 700 K, minor structural anomalies with no space group modification at ~475 and ~495 K were found. These anomalies were assigned to the phase-segregated state existing within a narrow temperature interval and manifested in the sharp phase transitions at the differential scanning calorimetry (DSC) and electrical resistivity data reported by *Galdino et al.* [20]. This result is in contrast to our previous study, in which no signs of the phase transitions were found using DSC and heat capacity studies [12]. Although the spin-state crossover provides the essential features for understanding the magnetic behavior of $Co_3BO_5$, its impact on the electronic properties and charge ordering is still unclear, as well as the exact nature of the local mechanisms responsible for these transitions.

This work aims to study the lattice dynamics and the electronic transitions in $Co_3BO_5$ through the thermal expansion, electrical resistivity, and magnetic susceptibility measurements in an extended temperature interval. The crystal structure and an evolution of the magnetic moment from 300 K up to 1000 K were studied using the x-ray powder diffraction and magnetic susceptibility measurements. The electronic properties up to 820 K were investigated using the electrical resistivity and thermopower measurements. A strong anisotropy of the thermal expansion of $Co_3BO_5$ was found, which is proposed to be due to the orientation of the rigid $BO_3$ triangles and the difference in the thermal expansion of the octahedral complexes $M3O_6$-$M1O_6$-$M3O_6$ and $M4O_6$-$M2O_6$-$M4O_6$. The evolution of the effective magnetic moment with temperature reflects the spin-state transformation of $Co^{3+}$ ions caused by the extensive thermal lattice expansion. The electrical resistivity rapidly decreases on heating. The activation energy follows the thermal expansion variation showing two anomalies. We conclude that $Co_3BO_5$ undergoes two gradual electronic transitions at ~500 and ~700 K, which are manifested in the heat capacity, lattice thermal expansion, and electrical resistivity measurements, highlighting a strong interrelation between lattice and electronic degrees of freedom.

## II. EXPERIMENTAL TECHNIQUES

Samples of $Co_3BO_5$ were synthesized by conventional high-temperature solid-state reaction. At the initial stage, the stoichiometric mixtures of $Co_3O_4$ (99.7%, metal sbasis, Alfa Aesar) and $H_3BO_3$ (99.99%, Alfa Aesar) were thoroughly blended in an agate mortar and annealed in air at a temperature of 1223 K in a corundum crucible for 24 h. The compound was ground and pressed into a pellet in the form of disks with a diameter of 15 mm. The pellets are heat-treated at 1223 K for 24 h (the speed of the heating is 50°/h) followed by a decrease in temperature to 1023 K at a rate of 5°/h, holding at set temperature for 12 h and subsequent quenching to room temperature at a rate of ≈ 250°/sec. The X-ray phase analysis has shown a resulting product contained 96.6% $Co_3BO_5$ and 3.4% residual $Co_3O_4$.

X-ray powder diffraction (XRPD) data were collected on a PANalytical X'Pert PRO diffractometer with copper monochromator on CuKα radiation. The measurements were performed over the diffraction angle range 10–80° 2θ in the temperature interval 300-1000 K using an Anton Paar HTK 1200N camera. The crystal lattice parameters were refined by the derivative difference minimization method [23].

The electrical resistivity was measured in the temperaturerange 450-820 K using an experimental set-up for thermopower and resistivity measurements [24]. The resistivity was measured with a standard four-point configuration. For the experiment, a $Co_3BO_5$ sample in the form of a regular parallelepiped of 14 mm in length and 0.9 mm in thickness was cut from a pellet.

For the magnetic measurements powder sample of $Co_3BO_5$ consisting of crushed single crystals [6], pressed as a pellet, so that the resulting powder grains were randomly oriented, was used. Two sets of measurements were performed: i) SQUID measurement on a 14.0 mg sample and applied field of 1 kOe, in the range $100 < T < 400$ K, and ii) in a VSM (vibrating sample magnetometer, with a peak amplitude of 2 mm and a $f$ = 40 Hz, $\Delta t$ = 5 seconds) placed in the oven option of a Quantum Design Physical Properties Measurement System (PPMS), on a 5.9 mg and an applied field of 10 kOe, in the $300 < T < 1000$ K range. In the higher $T$ range a special sample holder was used. Warming is achieved by using a heater patterned onto the sample holder. A thermocouple embedded on the back side of the sample holder measures the temperature at the sample region. We have used a dry mounting method, mounting the sample onto the heater stick using mechanical force. Sample and heater stick are wrapped with a copper-foil shield to keep the heat and minimize thermal gradients between the sample and the holder. Sample chamber is maintained in high vacuum (less than $10^{-5}$ torr) to avoid thermal contact leaks. A separate measurement of sample holder was performed at the same field and temperature routine, yielding a continuous diamagnetic contribution, which was added to the total magnetization to correct for any background effect.

**III. RESULTS**

Two representative XRPD patterns at 300 K and 1000 K are displayed in Fig. 2. The resulting crystal structure is orthorhombic with the space group *Pbam*, and no extra peaks were detected. Therefore, there is no structural phase transition up to 1000 K for the compound $Co_3BO_5$. The resulting lattice parameters are shown in Fig. 3 and listed in Table S1 of Supplementary Materials [25]. The comparison of the present data with the data available for single crystal [12] and powder sample [20] reveals the differences between refined room-temperature structural parameters of $Co_3BO_5$, which are probably due to the difference of the synthesis technique, as well as technological factors within the same synthetic technique.

An important conclusion is that all crystal structure data available up till now display the general peculiarities of the temperature behavior of the lattice parameters. Despite the sample state (single or poly-crystalline), the *a*-parameter shows a decrease, followed by an increase, upon warning. The *b*- and *c*- parameters show rapid and then slow growth in similar *T* intervals. Strongly nonlinear temperature behavior of lattice constants means that these changes are dominated not only by the phonons' variation, but that there are other considerable contributions.

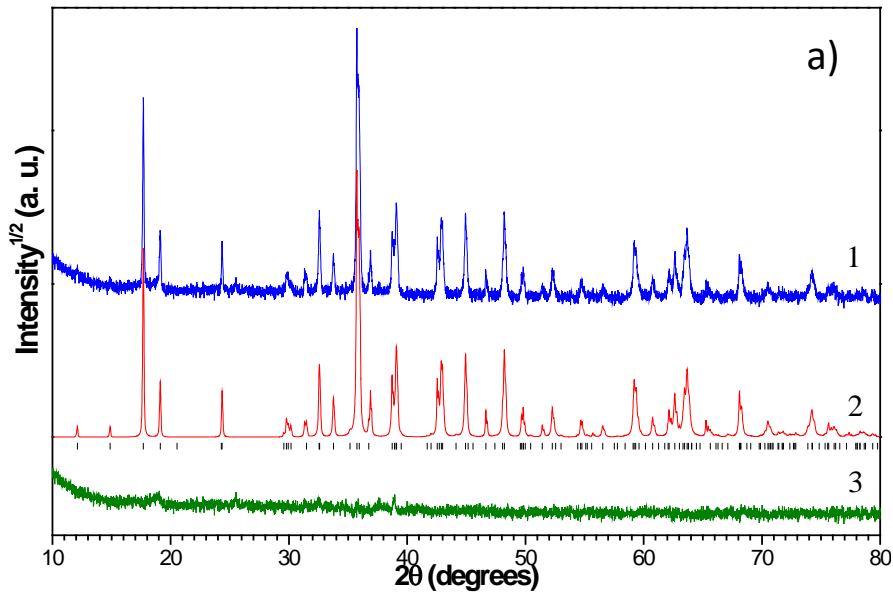

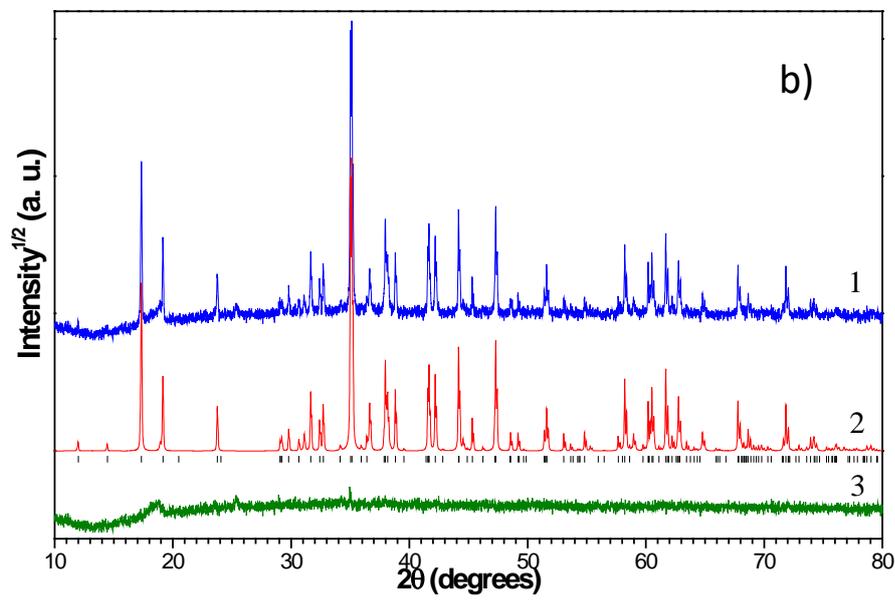

Fig. 2. Observed (1), calculated (2), and difference (3) XRPD patterns of $Co_3BO_5$ at 300 K (a) and 1000 K (b).

A thorough analysis has shown that the temperature interval of interest can be divided into two parts where lattice parameters demonstrate quite different behaviors. An inflection point at $T_C$=550 K is clearly observed at the temperature curves of *a*- and *c*- parameters. The same temperature separation can be found for the *b*-parameter, but it is not so noticeable. Below and above the critical temperature $T_C$, the thermal variations of lattice parameters can be fitted well by second-order polynomials with parameters collected in Table S2 [25].

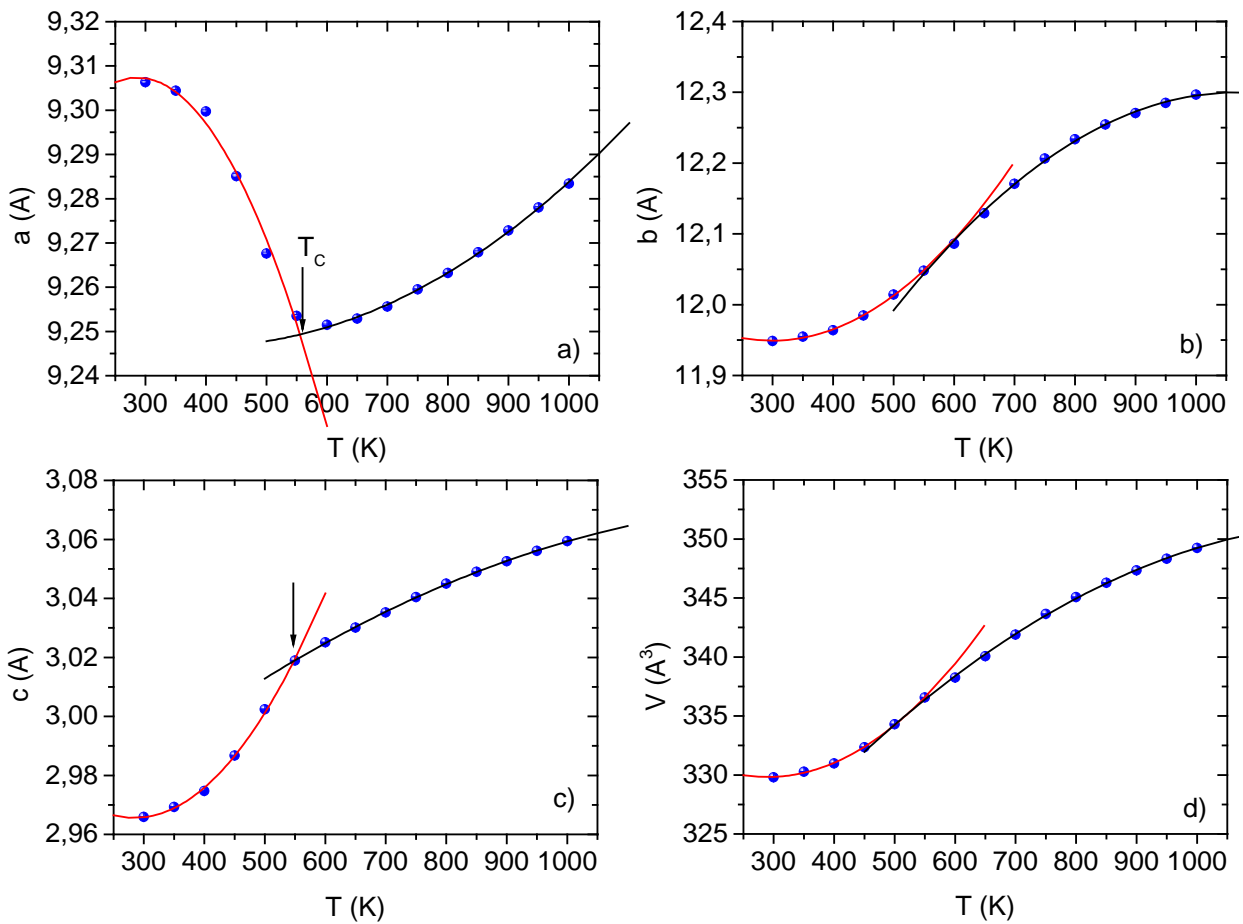

Fig. 3. (a)-(c) temperature dependences of the *a*-, *b*-, and *c*-lattice parameters and (d) - unit cell volume of $Co_3BO_5$ (symbols). Solid lines are fit curves below (red) and above (black) critical temperature $T_C$ (shown by an arrow). Unless shown, error bars are commensurate with the symbol size.

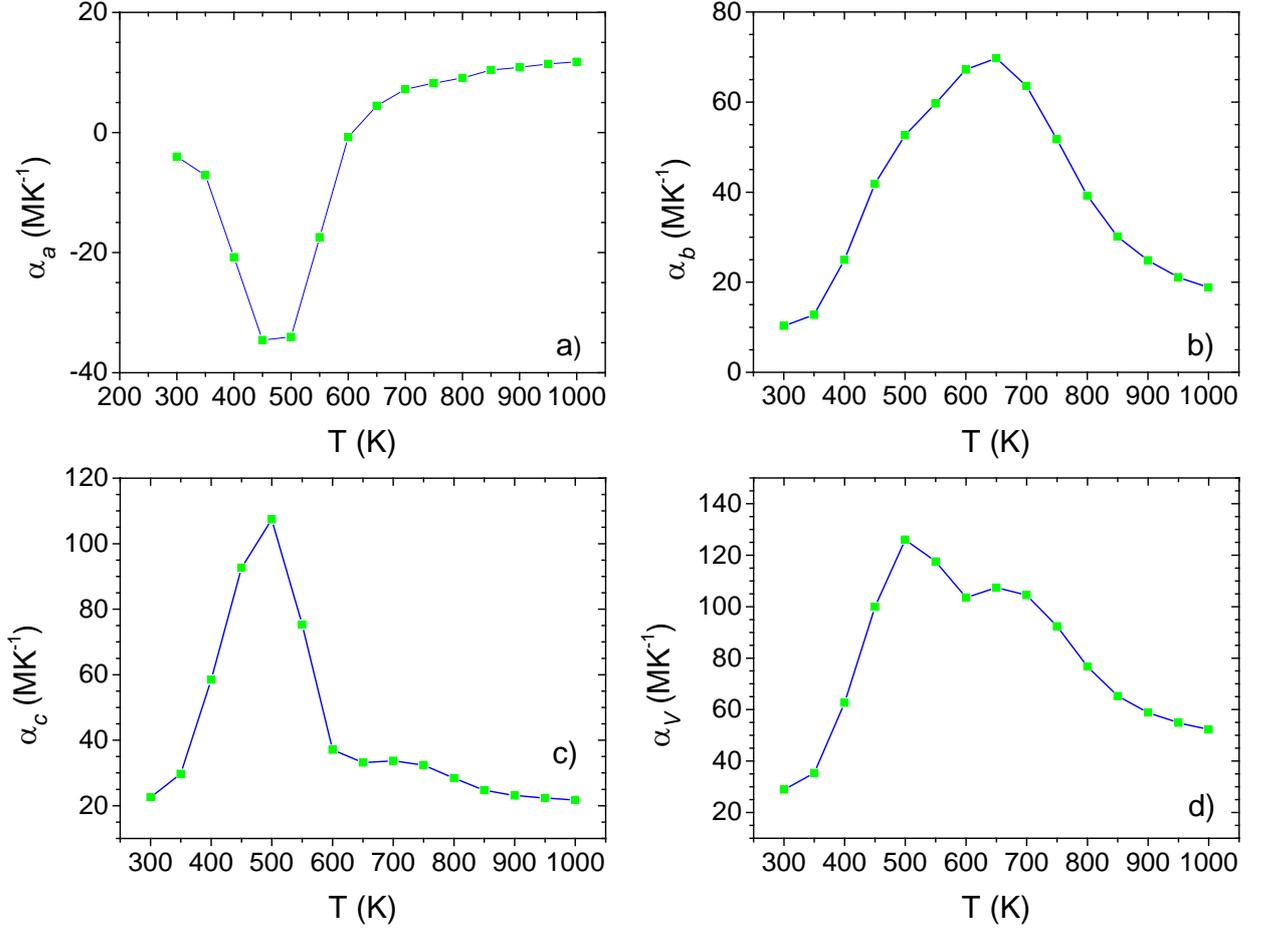

Fig. 4. (a)–(c) the thermal expansion coefficicents of *a*-, *b*-, and *c*- lattice parameters and (d) unit cell volume of $Co_3BO_5$ as a function of the temperature.

Figure 4 diplays the thermal expansion coefficients of the lattice parameters (*a*, *b*, *c*), and unit cell volume (*V*) expressed as following

$$\alpha_a = \frac{1}{a}\frac{da}{dT}, \quad \alpha_b = \frac{1}{b}\frac{db}{dT}, \quad \alpha_c = \frac{1}{c}\frac{dc}{dT}, \quad \text{and} \quad \alpha_V = \frac{1}{V}\frac{dV}{dT}. \tag{1}$$

The heating at the *T* range 300-550 K is accompanied by a sharp thermal expansion along the *c*-axis, manifesting in a broad maximum centered at 500 K with $\alpha_c \approx 110$ MK$^{-1}$ and by simultaneous compression along *a*-axis with negative coefficient $\alpha_a = -35$ MK$^{-1}$. Above 550 K the main contribution to the lattice expansion is given by *b*-parameter showing a broad maximum at 700 K ($\alpha_b \approx 70$ MK$^{-1}$). A fingerprint of this intensive lattice expansion is found in the $\alpha_a$ thermal behavior. One can conclude that the thermal expansion of the crystal lattice of $Co_3BO_5$ is strongly anisotropic. The cobalt ludwigite demonstrates the largest anisotropy of the thermal expansion among the parent compounds studied up till now, for example, iron borate $FeBO_3$ [26], hulsite $(Fe^{2+},Mg,Fe^{3+},Sn)_3(BO_3)O_2$ [27], iron vonsenite $(Fe,Mg)_3BO_5$ [27], and oxoborate $Fe_3O_2(BO_4)$ [28]. The next is that the volume thermal expansion demonstrates two well-resolved anomalies, which positions coincide with those observed in the temperature dependencies of the heat capacity [12] and the resistivity (see text below).

The electrical resistivity of the $Co_3BO_5$ single crystal at $T=200$ K is extremely large, ~$10^{10}$ Ohm·cm [18]. Therefore, the cobalt ludwigite is a robust insulator. However, the resistivity rapidly decreases with the heating, changing of about 7 orders of magnitude up to $T=400$ K (Fig. 5a). The resistivity data obtained for the polycrystalline sample in the more extended range $T=450$-$820$ K fit well to the general trend reported for the single crystal. The resistivity value at the highest measured temperature is ~$10^{-1}$ Ohm·cm.

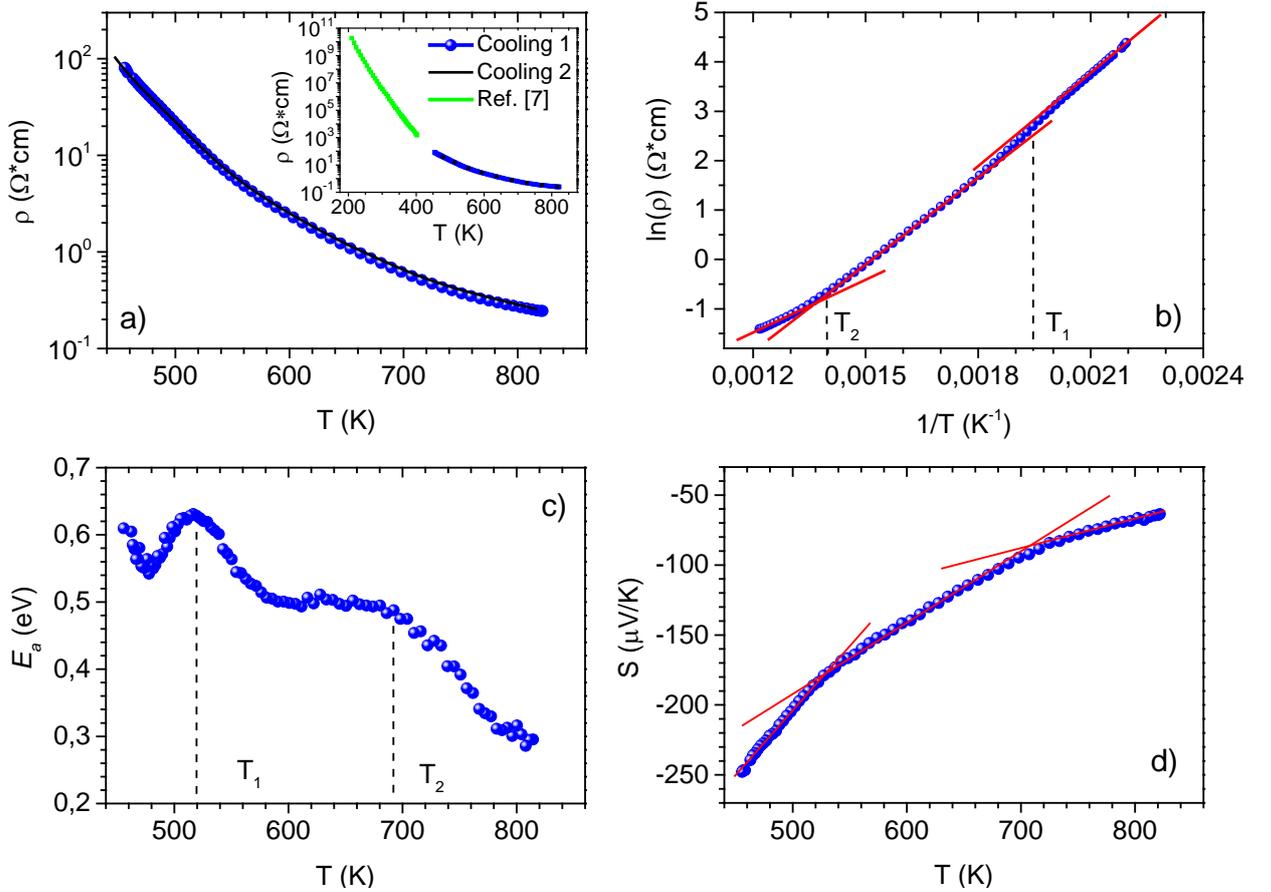

Fig. 5. a) Temperature dependence of the electrical resisitivity of $Co_3BO_5$. Four runs of measurements were performed, two - on heating and two on cooling. The inset: electrical resistivity as a function of the temperature measured for single crystal [18] and powder sample. b) The electrical resistivity presented in logarithmic scale as a function of the inverse temperature. The blue circles correspond to the data shown in (a), the red lines show the linear behaviors of Arrhenius plot. c) The temperature dependence of an activation energy. Dashed lines in (b) and (c) indicates the electronic transitions. d) Thermoelectric power as a function of the temperature.

At high temperatures, the thermo-activation conductivity mechanism is assumed to be prevalent [18], and the resistivity obeys satisfactorily the Arrhenius law demonstrating a linear relationship between $\ln\rho(T)$ and $1/T$ (Fig. 5b). Nevertheless, two inflection points are seen indicating the changes in the slope of $\ln\rho(T)$ and, hence, thermo-activation energy. These anomalies are better seen on the temperature dependence of the local activation energy $E_a = \frac{d(\ln \rho(T))}{d(k_B T)^{-1}}$ obtained by the direct differentiation of the experimental data $\rho(T)$ (Fig. 5c). The activation energy firstly decreases from $E_a = 0.61\pm0.02$ eV in the range $T$=450-480 K followed by the intensive maximum centered at $T_1$=520 K. The second anomaly appears near $T_2 \approx 700$ K, above which the activation energy rapidly decreases down to a value of $E_a = 0.28\pm0.02$ eV at 820 K. On the contrary to the former, the second electronic transition is broad and extended. In addition, the thermopower of $Co_3BO_5$ as a function of the temperature was measured (Fig. 5d). Seebeck coefficient is negative in the entire temperature range, indicating that the electrical conduction is intrinsic with high-mobility electrons and low-mobility holes.

We have measured the magnetization of the powder sample as a function of temperature. The linear dependence of the magnetization as a function of applied field for fixed temperature was verified, thereafter, we deduced the magnetic susceptibility as the ratio of measured magnetization versus applied field, thus

$$\chi(T) = M(T)/B \qquad (2)$$

In Fig. 6 we have depicted the data as $\chi^{-1}(T)$.

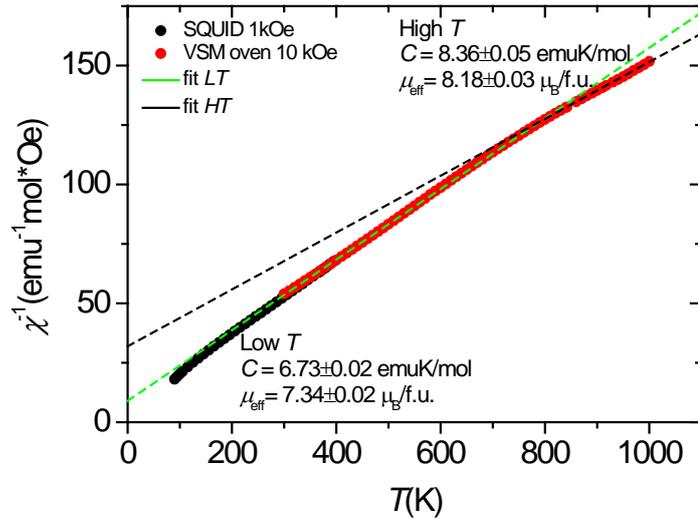

Fig. 6. Inverse of the magnetic susceptibility of $Co_3BO_5$ polycrystalline sample. SQUID measurement and VSM measurement in an oven are labeled by black and red symbols, respectively. The green and black dotted lines are the Curie Weiss law fits in the ranges of 300<$T$<550 K and 850<$T$<1000 K, respectively.

We note an excellent continuity of both data sets, i.e. the slope of the (i) data, extended to high temperature till 600 K, and the (ii) data coincide. Therefore, applying the Curie Weiss formula for a paramagnet

$$\chi^{-1}(T) = \frac{1}{C}(T - \theta) \qquad (3)$$

with $C = N\mu_{eff}^2/3k_B$ one obtains $\mu_{eff} = 7.34\pm0.02$ $\mu_B$/f.u. ($\mu_{eff} = 4.24$ $\mu_B$/Co) and $\theta = -60\pm2$ K in the temperature region 300 till 550 K. Note: In this analysis we do not consider a correction of TIP (Temperature Independent Paramagnetism) or $\chi_0$. We let any departure of the Curie law to be included in the dependence of the effective moment with temperature. The high $T$ susceptibility from 600 K to 1000 K changes slope continuously, reaching a nearly constant slope at higher temperatures with $\mu_{eff}$ =8.18±0.03 $\mu_B$/f.u. ($\mu_{eff}$ =4.73 $\mu_B$/Co) and $\theta$ = -267±8 K. To view this evolution more clearly we have depicted in Fig. 7 $\mu_{eff}$ *vs* $T$ as calculated from the tangent to the experimental curve $d(\chi^{-1})/dT$, and from the Curie Weiss law analysis.

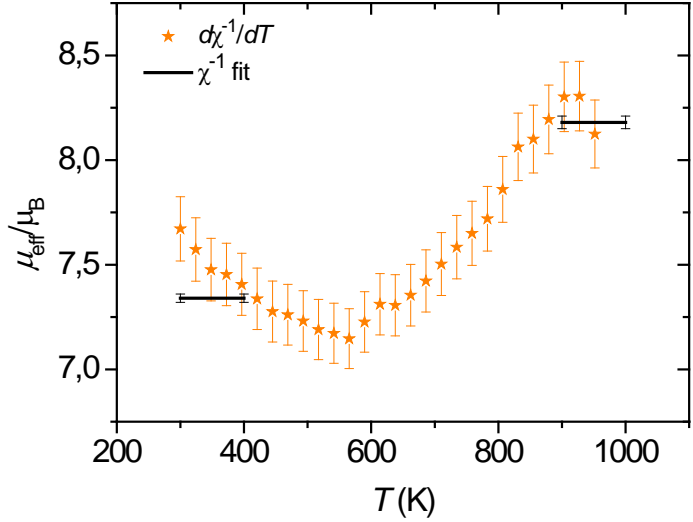

Fig. 7. The solid lines are $\mu_{eff}$/f.u. as deduced from the Curie Weiss law fit in two segments of the $\chi(T)^{-1}$ curve in Fig. 6. The symbols are $\mu_{eff}$/f.u. deduced from the tangent to the $\chi(T)^{-1}$ curve.

The value $\mu_{eff}$ =7.34±0.02 $\mu_B$/f.u. is somewhat higher than the previous value reported in Ref. [12] $\mu_{eff}$ =5.7±0.1 $\mu_B$/f.u. The difference stems from the absence of any $\chi_0$ correction of the data in this work. In this way we can compare those data with the new high temperature susceptibility data. At any rate, we observe a clear increase of $\mu_{eff}$ from a low value below 600 K to a high value above 800 K, corresponding to the spin state crossover reported earlier.

## IV. DISCUSSION

To understand the anisotropy of lattice thermal expansion observed in $Co_3BO_5$, it is necessary to turn to the crystal chemistry of borate compounds. The high-temperature crystal chemistry of borates predicts minimal thermal expansion in the plane of boron-oxygen triangles and maximal one perpendicular to them [29]. In the ludwigite, the $BO_3$ triangles are oriented parallel to the (001) plane, therefore the thermal expansion is expected to be minimal in this plane. Indeed, the expansion of the crystal structure of $Co_3BO_5$ is minimal along the *a*-axis and maximal along the *c*-axis, i.e. parallel and perpendicular, respectively, to the planes of the $BO_3$ groups. On heating, the configuration and size of the $BO_3$ and $BO_4$ polyhedra and rigid complexes consisting of these polyhedra are practically unchanged [29]. This is in agreement with the conclusions derived from the single-crystalline x-ray diffraction study of the $Co_3BO_5$, which revealed that in the temperature range $T$=300-700 K the individual bond-lengths B-O and bond-angles O-B-O within the trigonal group are not affected by the heating and their average values are kept constant [12].

In the ludwigite structure, three oxygen atoms O2, O3, and O5 (hereinafter we use the atom's numbering proposed in the work [12]) coordinate boron atom forming the trigonal group $BO_3$, while two remaining atoms O1 and O4 are independent oxygen atoms. It is expected that namely, the bonds toward the O1 and O4 atoms are most "flexible", that is, have more freedom for a change. From the six oxygen atoms coordinating the metal atoms at nonequivalent positions M1 and M3, four simultaneously belong to the boron-oxygen group (2xO2, 2xO3), while only two remaining atoms (2xO1) are independent oxygen atoms. For metal atoms occupying the M2 and M4 sites, this ratio is inverse: two oxygen atoms (O2 and O3) include in the boron coordination, while the remaining four oxygen atoms (2xO1 and 2xO4) only coordinate metal atoms. Already from these crystal-chemistry considerations, it follows that the octahedra $M2O_6$ and $M4O_6$ should be more "flexible" and have a greater potential for the stretching/compressing and, therefore, for deformation under external influence. The pliability of the bond-lengths of the $M2O_6$ and $M4O_6$ octahedra leads to their high ability to adapt to the ion size, whether this change is caused by the temperature, spin-state transition, or substitution [3,4,6-12].

According to the relative scale of "rigidity", reflecting the number of the dependent oxygen atoms involved into the boron atom coordination, the polyhedra can be divided into three groups: i) rigid polyhedra are the $BO_3$ triangles (100% bonds with boron atom), ii) relatively rigid-octahedra $M1O_6$ and $M3O_6$ (~ 70%), and iii) soft - octahedra $M2O_6$ and $M4O_6$ (~ 30%). While rigid triangles are isolated from each other, the second and third groups of polyhedra condense into complexes. Three vertex-linked octahedra form a linear chain ("triad" or "trimer") $M3O_6$-$M1O_6$-$M3O_6$ elongated in the $b$-axis. At the same time, the softer octahedra $M2O_6$ and $M4O_6$ are linked by the common edge (via two O4 atoms) and form the chain $M4O_6$-$M2O_6$-$M4O_6$ elongated in the direction close to $a$-axis (the angle relative to the $a$-axis is ~ 29.261 °). As it will be shown below, these two octahedral complexes demonstrate rather different responses to the lattice expansion.

The M1-O1-M3 bond-angle for the linear chain M3-M1-M3 is about 114° at 296 K and progressively increases up to ~120° at 703 K (Fig. 8a), displaying the maximal increment among all bond-angles in the structure. In such a chain if the octahedra have diverged relative to the common vertex O1, then the bond-angles increase, the chain is elongated and the structure expands. Such a linear chain is expected to have a maximum length at the bond-angle equal to 180°. The temperature-induced increase in the bond-angle within the $M3O_6$-$M1O_6$-$M3O_6$ octahedral complex correlates well with the experimentally observed thermal expansion along the *b*-axis in the same *T*-range (Fig. 4b). On the contrary, the M2-O4-M4 bond-angles within the polyanionic complex built of "soft" octahedra $M4O_6$-$M2O_6$-$M4O_6$ demonstrate an opposite effect and monotonically decrease with temperature (Fig. 8b). The decrease in the bond-angles ∠M2-O4-M4 takes place both between atoms along the chain, where the bond has an angle of about~90°, and between atoms belonging to adjacent chains with the angle of ~165°. The thermal expansion along such a chain is expected to be minimal, or even negative, in the case of a decrease in these angles, i.e., during compression of the chain.

The crystal structure analysis revealed that all bond-angles involving the M4 site (the angles Co4-O-Co*i*, *i*=1, 2, 3, 4) decrease with heating (Fig. 8c) [12]. This indicates the special position of the M4 site compared to other metal sites. Note, that the M4 site like the B site is located in the voids between the *bc*-layers built of the edge-sharing octahedra $M1O_6$, $M2O_6$, and $M3O_6$, and their role in connecting these layers along *a*-axis. The expansion/compressibility of the crystal structure in this direction is determined by the combination of softer bonds included in the $M4O_6$ octahedron and rigid bonds in the $BO_3$ group. This leads the structure first to compress due to a decrease in the bond-angle via M4 site. This compression occurs until the compression limit of the borate group is reached. After that, the expansion begins. This observation is firstly consistent with the view of $BO_3$ polyhedra as a rigid structural unit and, secondly, explains the negative thermal expansion along the *a*-axis if it arises from a deformation mechanism involving the decrease in the bond-angle via the M4 site (Fig. 4a). A number of materials display the negative thermal expansion occurring via the "bridging oxygens" mechanism. For the explanation of this effect, the RUM (Rigid Unit Modes) and QRUM (Quasi Rigid Unit Modes) models were used [30, 31]. Detailed analysis is needed to find out whether the ludwigite structure can support RUMs and, if so, of what type.

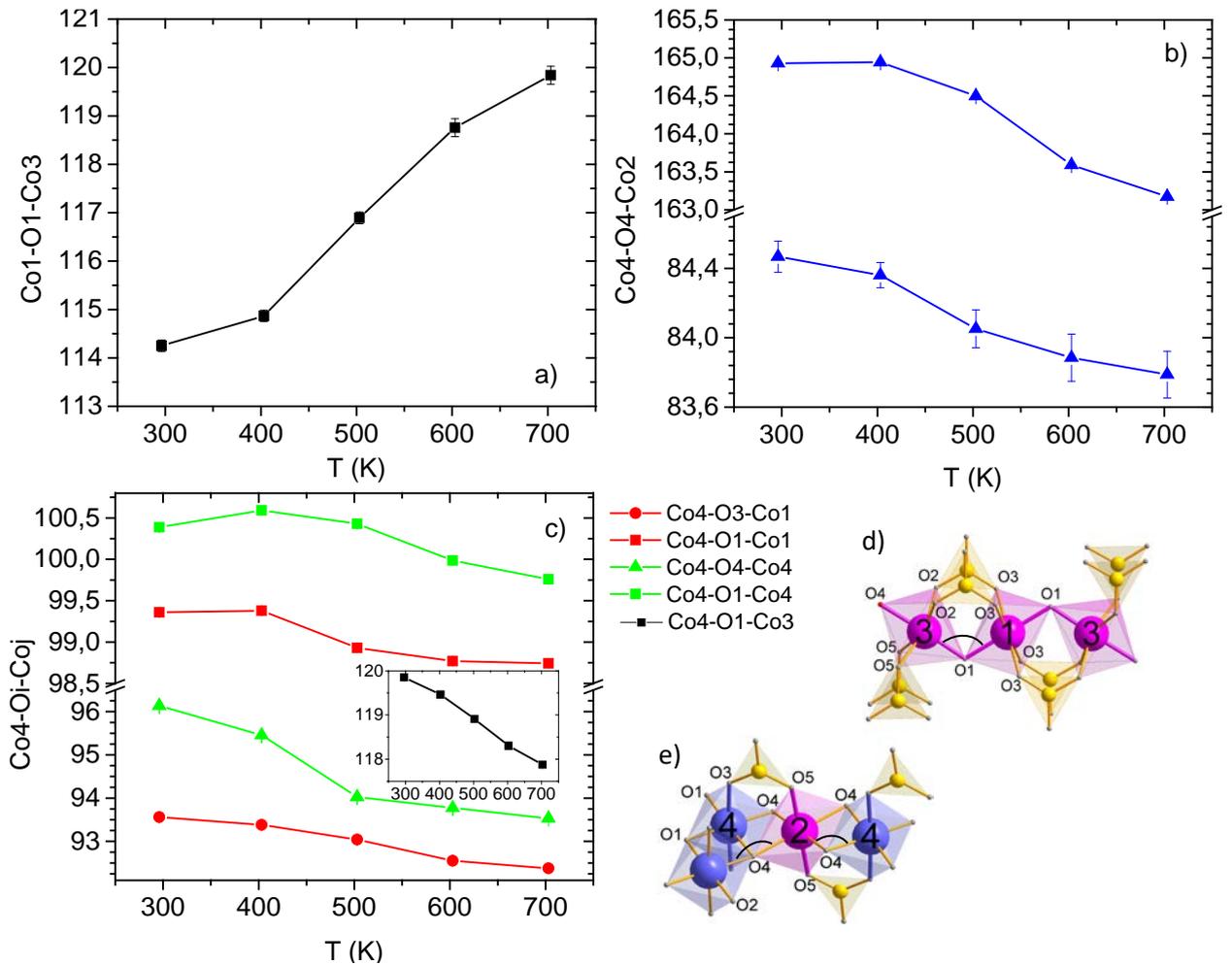

Fig. 8. The bond-angles between (a) Co1 and Co3, (b) Co2 and Co4 atoms inside the $Co3O_6$-$Co1O_6$-$Co3O_6$ and $Co4O_6$-$Co2O_6$-$Co4O_6$ octahedral complexes. c) the bond-angles between Co4 atom and Co1, Co3, and Co4 atoms. d) and e) representations of the above bond-angles.

The thermal deformations reflect mainly the change in the vibration atomic size. Such changes can also be induced by isomorphic substitution under constant thermodynamic parameters. The first case corresponds to the thermal expansion, and the second to the chemical/composition expansion. The chemical expansion in Co-containing ludwigites was studied in Ref.[12] and it was found that the lattice parameters and unit cell volume are determined predominantly by the size of the metal ion located at the M4 site (Fig.14 [12]). By developing this idea, we have synthesized the solid solutions of $Co_{3-x}Fe_xBO_5$ (0.0≤$x$≤1.0) with gradual substitution of $Co^{3+}$ ion ($r_i$=0.545 Å for the LS state [22]) by $Fe^{3+}$ (0.645 Å) at the M4 site. Figure 9 shows the variation of the lattice parameters and unit cell volume for the solid solutions as a function of the effective size of $M4O_6$ octahedron, expressed in terms of the mean bond-length <M4-O>. The similarity of the increment in the $b$-, $c$- lattice parameters suggests that the Fe-substitution acts as the evolution of temperature does. More importantly, the $a$-parameter demonstrates a strong sensitivity to the sort of metal ion at the M4 site. The negative expansion of this parameter for $Co_3BO_5$ caused by the temperature is reversed, i.e. becomes positive in the case of chemical expansion caused by the substitution. This feature suggests that anomalous thermal expansion along *a*-axis can be tuned by the M4 site substitution. So, by substituting the atoms at the given site, one can change the magnitude and sign of the thermal expansion and, probably, achieve the zero thermal expansion or expansion in a narrow temperature range, thereby approaching the practical application. The zero thermal expansion effect is known to have large technical applications such as high-precision instruments, protection shields for inflammable and explosives, and catalyst supports [32-34].

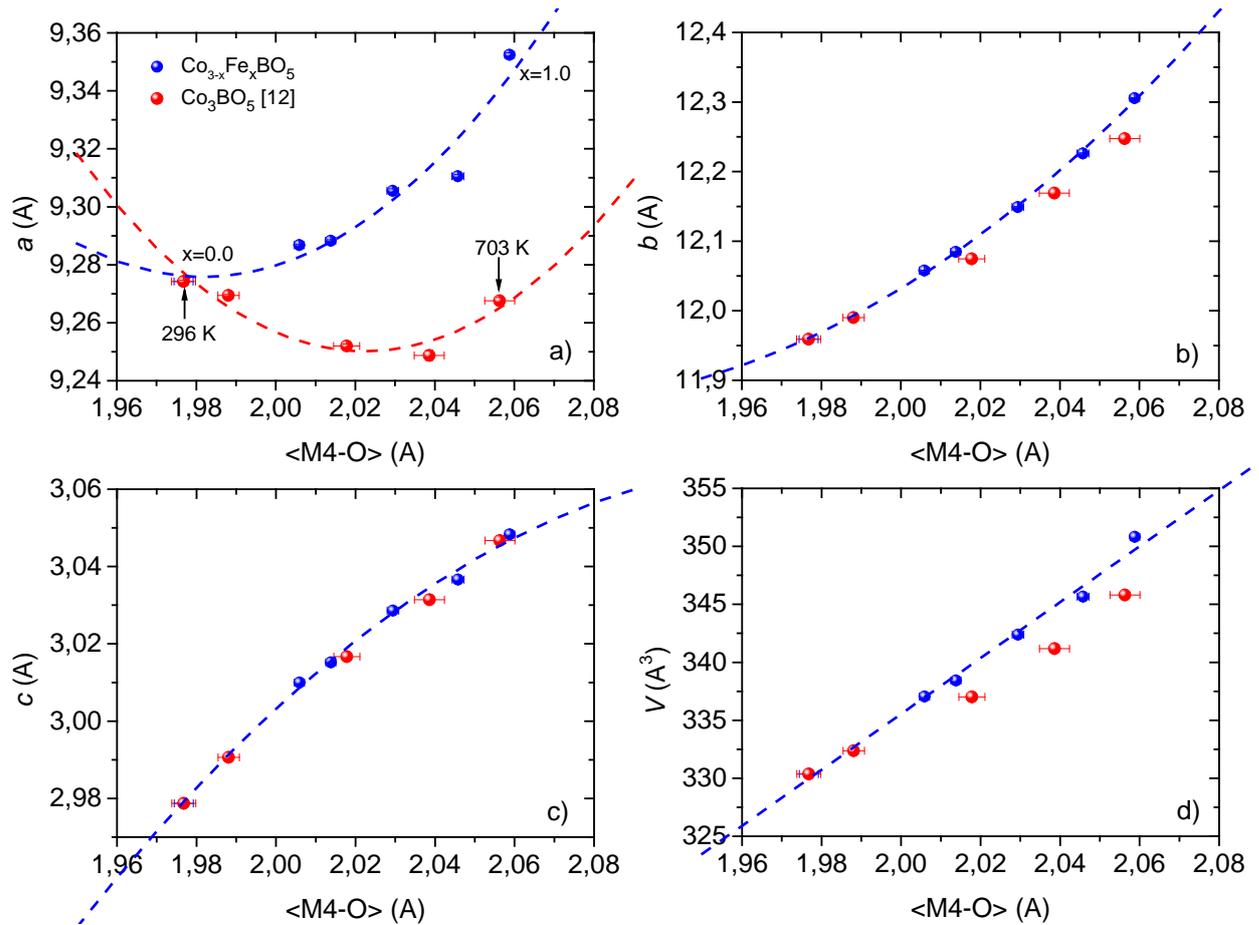

Fig. 9 (a)–(c) the variation of *a*, *b*, and *c* lattice parameters and (d) unit cell volume presented as a function of the size of $M4O_6$ octahedron for $Co_{3-x}Fe_xBO_5$ solid solutions ($0.0 \leq x \leq 1.0$) on the substitution of the $Fe^{3+}$ for $Co^{3+}$ (blue circles) and for $Co_3BO_5$ on heating (red circles) [12]. The dashed lines are the guides for the eye.

The theoretical band calculations of the $Co_3BO_5$ revealed that the charge-ordering is not sensitive either to the spin state of the $Co^{3+}$ ion or to the magnetic structure (ferri- of ferro-), but is related to the features of the crystal structure. The latter can be understood in the following way. The boron atoms pull oxygen atoms away from metal atoms, causing the elongation of the corresponding bond-lengths. The more oxygen atoms simultaneously participate in the coordination of boron, the longer the average bond length of the octahedron and, consequently, the lower the effective electrostatic field acting on the metal ion. This leads to the predominant filling of these sites with ions with a lower charge state. Following this approach, the M1 and M3 sites possessing the most number of oxygen bonds with boron should show the greatest preference for sheltering the divalent ions. On the contrary, the M4 site possessing the smallest number of oxygen bonds with boron is prone to adopt a +3 charge state. This type of reasoning fully agrees with actual experimental charge states at metal sites observed in the ludwigites with $Me^{2+}$=Co, Fe, Mn, Cu, Ni, Mg, and $Me^{3+}$=Co, Fe, Mn, Ga, Al, etc.. [3,4,11,35-38].

The rigidity of the BO$_3$ triangle units requires their primary ordering over the crystal, which leads to the build-up of a rigid triangular sublattice. The octahedral sublattice consisting of softer polyhedra adjusts to the triangular one. If the number of divalent and trivalent cations per formula unit is equal, for ludwigites this ratio $q = \frac{M^{3+}+B^{3+}}{2 \cdot M^{2+}} = 1$ and for pyroborates $q = \frac{2 \cdot B^{3+}}{2 \cdot M^{2+}} = 1$, then a condition arises for the layered ordering of these cations and, as a consequence, the layered charge ordering. This leads to the M2 site, adjacent to the M3 sites, but having a different number of oxygen bonds with boron, that demonstrates a charge state close to +2. Thus, the Co$_3$BO$_5$ can be represented as a layered material composed of 2D infinite [Co$^{2+}$O$_6$]$_\infty$ layers along the *bc* plane. The layers are connected along *a*-axis *via* [BO$_3$] triangles and [Co$^{3+}$O$_6$] octahedra at the M4(4h) site, which are alternatively distributed. Because the Co$^{3+}$ ion is locked in the site between the boron atoms, it is somewhat subject to the harsh bonding conditions inherent in the borate group. Perhaps, this is a reason why the trivalent cobalt ion is forced to be in a low-spin state at low temperature. As the temperature increases, the bond-lengths and bond-angles increase leading to the lowering of the crystal field magnitude (10*Dq*) at metal sites. As a result, the Co$^{3+}$ ion undergoes the smooth low-spin to a high-spin state transition. We want to emphasize that the spin-state transition is not a sharp phase transition, but a continuous process with smooth growth of the high-spin term concentration $n_{HS}$ with temperature similar to LaCoO$_3$ [39]. Even for LaCoO$_3$ where this process is the most intensive at *T*~100 K following the maximum of the magnetic susceptibility, the average magnetic moment at *T*=1000 K reaches the value 1.9 $\mu_B$ (for HS it is expected to be equal 2) [40]. So, the theoretical result from [12] should be understood that HS configuration is dominant, but it hardly reaches the 100% at *T*=700 K.

The tendency of layer-by-layer charge ordering is well traced to the example of the Mn-Mg-B-O system, within which the structural series of warwickite-orthopinakiolite-hulsite-ludwigite has been successfully synthesized [41]. With the Mg concentration increase, the tendency for substitutional atoms to form the (MgO$_6$)$_\infty$ layers is clearly seen, while trivalent manganese ions fill sites in the voids between these layers, like the boron atoms.

The next striking feature of the ludwigites is partial charge orderings, which involve not all, but only specific metal sites. The M3-M1-M3 octahedral complex does not participate in charge-ordering and, in fact, has high structural and electronic stability. The antiferromagnetic interaction between the magnetic moments of $Me^{2+}$=Co, Fe, Mn, Cu belonging to this complex turns out to be the strongest leading to uncompensated magnetic moment observed in most of the ludwigites studied to date [4,6-11,42,43]. Moreover, the ferrimagnetic ordering of $Co^{2+}$ magnetic moments at M1 and M3 sites in $Co_2AlBO_5$ is accompanied by the jump-wise change in the lattice parameters as it was found using the temperature-dependent x-ray diffraction measurements [42] and reflects a strong interrelation between the spin and lattice subsystems belonging to this octahedral complex. In $Fe_3BO_5$ a partial charge-ordering with the formation of dimeric states is accompanied by an isostructural phase transition *Pbam*(No. 55)→*Pbam*(No. 62) and doubling of the *c*-lattice parameter ($T_{CO}$= 283 K). The experimental confirmations of such structural transformation were found using XRD and NPD studies [14-17]. A broad minimum in the resistivity along the *c*-axis was observed, which could be attributed to the charge-ordering but, however, has not yet found reliable confirmation [44].

Nevertheless, no signs of the superstructure implying a partial charge-ordering in $Co_3BO_5$ were found. This may indicate that charge-ordering transition either does not occur or occurs but without the accompanying structural phase transition. Indeed, the gradual charge-ordering transition does not require symmetry changing, unlike a phase transition. If such transition involves all metal sites then this electron exchange process through charge delocalization can be formulated as $2 \cdot Co_{HS}^{2+} + 1 \cdot Co_{HS}^{3+} \rightarrow 3 \cdot Co^{2.33+}$ and is manifested as a tendency to converge to the same valence state. Experimentally, this electronic process is expressed in the gradual equalization of the average size of Co ions on heating (Ref. 12, Fig. 12a). If the charge transfer takes place through such a process, then charge carriers are created and it should be observable via the change of the electrical conductivity. This assumption is in agreement with the *n*-type conductivity deduced from our thermopower measurements (Fig. 5d).

According to DFT calculations of band structure, the $Co_3BO_5$ ground state is an insulator with a band-gap of 1.4 eV [12]. The disruption of the charge-ordering and the metallic properties are predicted at high temperatures. Note, that the GGA+U calculations can describe the initial and final states of this smooth process, but not the temperature-dependent intermediate states. The experimental observation of the large resistivity (Fig.5a) and the experimental value $E_g=2E_a$=1.7 eV at low temperatures [18] are in good agreement with the theoretical calculations in the low-temperature phase. Although in our experiment the metallic state has not been reached, a significant reduction of the activation energy from $E_a$=0.61±0.02 eV at 450 K to 0.28±0.02 eV at 820 K has been found, indicating the system undergoes a semiconductor-semiconductor transition (Fig.5c). The important finding is that the energy gap is strongly temperature-dependent and is related to anomalous thermal expansion. In fact, the energy gap follows the volume thermal expansion over the entire temperature range (Fig.10). Thus, the strong interrelation between the lattice and electron degrees of freedom inherent in ludwigites which is manifested in the $Fe_3BO_5$ as the related electronic and structural transitions at $T_{CO}$, here, in the $Co_3BO_5$, is manifested in the high sensitivity of the electronic subsystem to the thermal expansion.

The observed increase of $\mu_{eff}$ with temperature seems to correlate well with a maximum of the thermal expansion coefficient found at $T_1$ =500 K (Fig. 4c) on one hand, and the change of slope at ~700 K with the change of electric resistivity slope at the second electronic transition detected at $T_2$ =700 K on the same sample (Fig. 5c), on the other. At any rate, it is clear from the magnetic measurements that the electronic transitions detected in the electric resistivity measurements as a function of temperature also have a bearing on the average magnetic moment dependence on $T$ (Fig.10). In the previous study [12] it was shown that the average effective magnetic moment of $Co_3BO_5$ in the paramagnetic regime at low temperature $T<300$ K is compatible with two $Co^{2+}$ ions in the high spin state and orbital contribution, and the $Co^{3+}$ in the low spin state. The observed increase in magnetic moment, therefore, is consistent with the sensitivity of the Co4 octahedral site with temperature and the increase in the octahedral Co-O distances and the local distortion caused by the an evolution of the $Co^{3+}$ towards spin crossover to a high spin state.

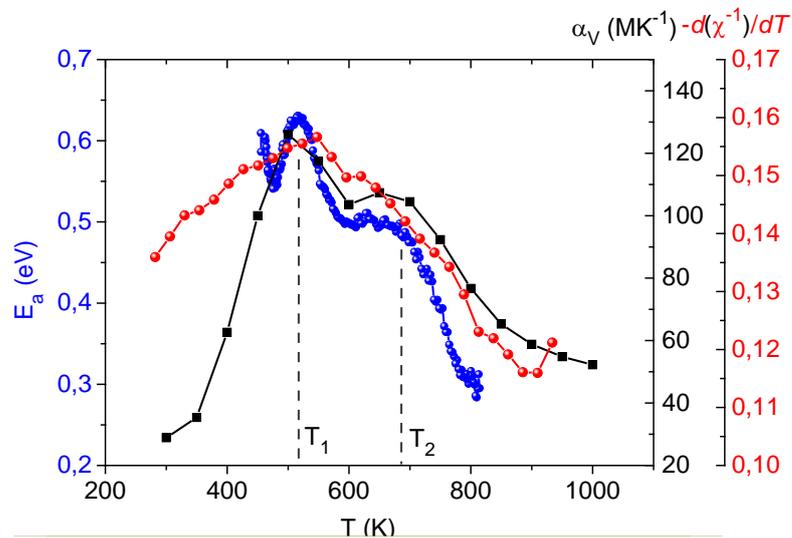

Fig. 10. Comparison of the volume thermal expansion coefficient and thermoactivation energy with the slope of the $-\chi^{-1}(T)$ curve. The dotted lines show the onset of electronic transitions at ~500 and 700 K.

It should be mentioned at this point that the present value $\mu_{eff}$/Co =4.24 and 4.72 $\mu_B$, for the low and high temperature phases, respectively, are not far from $\mu_{eff}$/Co =4.87 $\mu_B$ at 700 K, reported for the same nominal sample [20]. As was explained earlier [12], the magnetic moments, transition temperatures, lattice parameters of our sample and theirs differ a little. It may reflect the influence of the synthesis technique, as well as technological factors within the same synthetic technique. Besides, the cationic distribution and the different degree of ordering due to different thermal history may also be affected.

## V. CONCLUSION

We have studied a $Co_3BO_5$ by x-ray powder diffraction and electrical resistivity measurements at high temperatures. It was found that the structural symmetry remains with the orthorhombic one in the studied temperature range from 300 to 1000 K, and no structural phase transitions were detected. The deduced temperature-dependent lattice parameters displayed a strong anisotropy of the thermal expansion: on heating, the crystal structure expands along *b*-, and especially along *c*- axes, and simultaneously contracted along *a*-axis ($T<T_C$=550 K). It can be explained by the orientation of the rigid $BO_3$ triangles and the different contributions of the octahedral complexes $M3O_6$-$M1O_6$-$M3O_6$ and $M4O_6$-$M2O_6$-$M4O_6$ to the thermal expansion. The decrease of the bond-angles involving the M4 site correlates with the negative thermal expansion along *a*-axis, assuming the crucial role of the given metal site in the thermal expansion of the ludwigites. The electrical resistivity measurements performed in the range of $T$=450-820 K revealed that the resistivity rapidly decreases on heating. The local activation energy $E_a$ follows the thermal expansion discovering two electronic transitions at ~500 and ~700 K whose positions coincide with features in the heat capacity [12]. The thermopower measurements revealed that the electron transport is the dominating conduction mechanism over temperature range. The observation of the electronic transitions correlate with the temperature-induced change in the effective magnetic moment associated with change in the spin state of $Co^{3+}$ ions. We conclude that with the temperature increase, the lattice parameters are changed not only due to conventional thermal expansion, but also due to the electronic processes associated with the spin-state and charge-ordering transitions. A strong interrelation of the crystal structure and electronic properties was found.

### ACKNOWLEDGMENTS

We are grateful to the Russian Foundation for Basic Research (project no. 20-02-00559, 21-52-12033) for supporting this paper. This work was performed within the framework of the budget project № 0287-2021-0013 for Institute of Chemistry and Chemical Technology SB RAS. We acknowledge financial support from the Spanish Ministry of Economy, Industry and Competitiviness (MINECO), Grant No. MAT2017-83468-R) and from the regional Government of Aragón (E12-20R RASMIA project)